\pdfoutput=1
\RequirePackage{ifpdf}
\ifpdf 
\documentclass[pdftex]{sigma}
\else
\documentclass{sigma}
\fi

\newcommand{\ovl}{\overline}
\newcommand{\tdef}{\text{def}}
\newcommand{\tcrit}{\text{crit}}
\newcommand{\llg}{\langle}
\newcommand{\rrg}{\rangle}

\newcommand{\RR}{\mathbb{R}}
\newcommand{\ZZ}{\mathbb{Z}}
\newcommand{\mA}{\mathcal{A}}
\newcommand{\mE}{\mathcal{E}}
\newcommand{\mH}{\mathcal{H}}
\newcommand{\mO}{\mathcal{O}}
\newcommand{\tH}{\tilde{H}}
\newcommand{\hf}{\hat{f}}
\newcommand{\hx}{\hat{x}}
\newcommand{\hH}{\hat{H}}
\newcommand{\oH}{\ovl{H}}
\newcommand{\dx}{\dot{x}}
\newcommand{\ddx}{\ddot{x}}
\newcommand{\dy}{\dot{y}}

\newcommand{\ga}{\gamma}
\newcommand{\De}{\Delta}
\newcommand{\ep}{\epsilon}

\newcommand{\Om}{\Omega}
\newcommand{\sg}{\sigma}
\newcommand{\Sg}{\Sigma}
\newcommand{\la}{\lambda}
\newcommand{\La}{\Lambda}
\newcommand{\p}{\partial}

\newcommand{\half}{\frac{1}{2}}

\newtheorem{Theorem}{Theorem}
\theoremstyle{definition}
\newtheorem{condition}{Condition}
\newtheorem{problem}{Problem}

\DeclareMathOperator{\prob}{Prob}

\begin{document}

\renewcommand{\thefootnote}{}

\renewcommand{\PaperNumber}{016}

\FirstPageHeading

\ShortArticleName{Some Open Problems in Random Matrix Theory and the Theory of Integrable Systems.~II}

\ArticleName{Some Open Problems in Random Matrix Theory\\ and the Theory of Integrable Systems.~II\footnote{This paper is a~contribution to the Special Issue on Asymptotics and Universality in Random Matrices, Random Growth Processes, Integrable Systems and Statistical Physics in honor of Percy Deift and Craig Tracy. The full collection is available at \href{http://www.emis.de/journals/SIGMA/Deift-Tracy.html}{http://www.emis.de/journals/SIGMA/Deift-Tracy.html}}}

\Author{Percy DEIFT}

\AuthorNameForHeading{P.~Deift}

\Address{Department of Mathematics, New York University, 251 Mercer Str., New York, NY 10012, USA}
\Email{\href{mailto:deift@cims.nyu.edu}{deift@cims.nyu.edu}}

\ArticleDates{Received October 11, 2016, in f\/inal form March 10, 2017; Published online March 14, 2017}

\Abstract{We describe a list of open problems in random matrix theory and the theo\-ry of integrable systems that was presented at the conference
\textit{Asymptotics in Integrable Systems, Random Matrices and Random Processes and Universality}, Centre de Recherches Math\'ematiques, Montr\'eal, June 7--11, 2015. We also describe progress that has been made on problems in an earlier list presented by the author on the occasion of his 60$^{\rm th}$ birthday in 2005 (see [Deift P., \textit{Contemp. Math.}, Vol.~458, Amer. Math. Soc., Providence, RI, 2008, 419--430, arXiv:0712.0849]).}


\Classification{37K15; 15B52; 34M55; 35Q53; 60F05; 60K35; 62H25; 82B44}

\renewcommand{\thefootnote}{\arabic{footnote}}
\setcounter{footnote}{0}

\medskip

At my 60$^{\rm th}$ birthday conference in 2005, I was asked to present a list of unsolved prob\-lems~\cite{PercyCM}. At my 70$^{\rm th}$ birthday conference in 2015, I~was asked to update the list, and also present any other problems that I found interesting and important. What follows is, more or less, the list of updates and problems that I presented. As in~\cite{PercyCM} the list is \begin{quotation}``written in an informal style. Detailed references are readily available on the web. In the text, various authors are mentioned by name: This is to aid the reader in researching the problem at hand, and is not meant in any way to be a detailed, historical account of the development of the f\/ield. I ask the reader for his/her indulgence on this score. In addition, the list of problems is not meant to be complete or def\/initive: It is simply an (unordered) collection of problems that I~think are important and interesting, and which I would very much like to see solved''.
\end{quotation}

I will proceed in the following way. For any given problem listed in~\cite{PercyCM}, I will f\/irst discuss the progress that has been made since 2005, and then I will discuss some aspects of the problems that remain open. The motivation for these problems and their signif\/icance will be discussed only brief\/ly: For more information, the reader should consult~\cite{PercyCM}. I will also discuss some open problems not on the previous list.

\begin{problem}[KdV with almost periodic initial data]\label{prob1}
Consider the Korteweg--de~Vries (KdV) equation
\begin{gather}\label{eq1}
u_t + u u_x + u_{xxx} =0
\end{gather}
with initial data
\begin{gather*}
u(x, t=0) = u_0(x),\qquad x\in \RR.
\end{gather*}
In the 1970's, McKean and Trubowitz \cite{McTr} proved the remarkable result that if the initial data~$u_0$ is periodic, $u_0(x+p)= u_0(x)$ for some $p>0$, then the solution $u(x,t)$ of \eqref{eq1} is almost periodic in time. This result leads to the following natural \textit{conjecture}: The same is true if $u_0(x)$ is almost periodic, i.e., if the initial data is almost periodic in space, the solution evolves almost periodically in time. This is a very challenging problem with many new and unique features. For example, for~$u_0(x)$ almost periodic, one does not know, by standard PDE methods, whether a solution $u(x,t)$ exists, even for a small time, never mind globally. In spectacular and seminal work over the last few years, D.~Damanik and M.~Goldstein \cite{DaGo1, DaGo2}, joined later by I.~Binder and M.~Lukic \cite{BiLu}, have partially resolved this conjecture in the af\/f\/irmative, namely, for small quasi-periodic analytic initial data with Diophantine frequencies, unique global solutions exist and are almost periodic in time. The key insight of Damanik et al.\ was how to utilize the formal integrability of KdV in an ef\/fective analytical manner. Not only are the arguments in Damanik et al.\ applicable to other (formally) integrable systems (for example~\cite{BiDaLuVa} proves a Toda lattice analog of the KdV theorem), but some of their arguments are also applicable to non-integrable systems. It is a natural, open and very challenging \textit{conjecture} that the result of Damanik et al.\ remains true for (suitable) perturbations of KdV, i.e., for (suitable) perturbations of KdV, the Cauchy problem with (suitable) almost periodic initial data~$u_0(x)$, has a unique global solution~$u(x,t)$, which evolves almost periodically in time. After all, the f\/inite-dimensional analog of this result is precisely the content of classical KAM theory. We note that such results for perturbations of KdV and the nonlinear Schr\"{o}dinger equation (NLS) with periodic initial data~$u_0(x)$, have already been proved by S.~Kuksin, T.~Kappeler and others.

With the discovery of quasi-crystals, one can anticipate, in particular, that interest in PDE problems on quasi-periodic backgrounds, will surely grow.
\end{problem}

\begin{problem}[universality for random matrix theory (RMT)]\label{prob2}
The question of universality for the distribution of the eigenvalues of a random matrix on the microscopic scale, goes back to the earliest days of RMT in the 1950s and the work of E.~Wigner, F.~Dyson and M.~Mehta, in particular. By 2005, universality for unitary invariant ensembles had been established, both in the bulk and at the edge, for very general analytic matrix distributions. Universality had also been established for invariant orthogonal and symplectic ensembles, both in the bulk and at the edge, for distributions of the form $e^{-\operatorname{tr} V(M)} dM$, where the matrices $M$ are $n\times n$ and
\begin{gather}\label{eq3}
V(x) = \ga_k x^{2k} + \ga_{k-1} x^{2k-1} + \dots + \ga_0, \qquad \ga_k>0
\end{gather}
is f\/ixed. For such distributions, the underlying equilibrium measure is supported on a single interval. However, in the important case with distribution $e^{-n \operatorname{tr} V(M)} dM$ where $V(x)$ is again f\/ixed as in~\eqref{eq3}, and where the underlying equilibrium measure for the problem can now be supported on more than one interval, a challenging technical issue for orthogonal and symplectic invariant ensembles involving a certain determinant (see $D_n$ in \cite{PercyCM}) remained unresolved. This issue was subsequently resolved in 2010--2011 by T.~Kriecherbauer and M.~Shcherbina \cite{KrSh} and M.~Shcherbina~\cite{Sh}.
\end{problem}

The challenge par excellence that remained for RMT as of 2005, was to prove universality in the bulk for general Wigner ensembles for which the entries $M_{ij}$ of the matrices are independent random variables, subject only to the given symmetry of the matrices. For such ensembles, the algebraic simplif\/ications and powerful analytical techniques (orthogonal polynomials, Riemann--Hilbert/steepest-descent methods) developed for invariant ensembles, are no longer applicable. In 1999, A.~Soshnikov \cite{Sosh} proved universality at the edge for Wigner ensembles. Soshnikov's proof utilized the so-called moment method introduced by E.~Wigner in the 1950's, together with certain trace estimates obtained earlier by Ya.~Sinai and A.~Soshnikov~\cite{SinaiSosh}. Apart from the technical, combinatorial virtuosity required to carry through his proof, Soshnikov introduced the strategy that foreshadowed all later work on universality for Wigner ensembles. Rather than computing explicitly the distribution of the eigenvalues at the top of the spectrum as $n=\dim M \to \infty$, as one does for invariant ensembles, Soshnikov just showed that the limiting distribution, whatever it was for any given Wigner ensemble, depended \textit{only} on the f\/irst two moments of the distribution of the (independent) entries $M_{ij}$ of the matrices~$M$. So in order to see explicitly what the limiting eigenvalue distribution was, all one had to do was to match up the distribution of the $M_{ij}$'s with a Gaussian distribution with the same two moments. For such Gaussian ensembles the limiting distributions had famously been computed explicitly by C.~Tracy and H.~Widom in 1994~\cite{TW} in terms of Painlev\'e functions. In this way universality at the edge was proved, by inference rather than by direct computation. This strategy is of course similar in spirit to standard proofs of the classical central limit theorem for the sum $S_n= x_1 + \dots + x_n$ of iid variables $\{x_i\}$ with distribution~$\chi$, where one shows that the sum of the variables, suitably scaled, converges to a~limiting distribution which only depends on the f\/irst two moments of~$\chi$.

The f\/irst proof of universality for eigenvalues in the bulk of the spectrum for a class of matrices broader than the class of invariant matrices, was given by K.~Johansson in 2001 \cite{JohUniv}. Johansson considered a Wigner ensemble of random matrices of the form $H_t = \frac{1}{ \sqrt{n}} \big(H + \sqrt{t} V\big)$, $t>0$. Here Hermitian $H$ is taken from a given Wigner ensemble of $n\times n$ Hermitian matrices with mean zero and variance~$\frac{1}{4}$, and $V$ is an $n \times n$ GUE matrix, also with mean zero and variance~$\frac{1}{4}$. For such ensembles, there are explicit formulae for the eigenvalue correlation functions closely related to analogous formulae for invariant Hermitian ensembles, and for f\/ixed $t>0$, Johansson was able to use these explicit formulae to prove universality as $n\to \infty$ for the eigenvalues of such matrices $H_t$ in the bulk by direct computation. En route to his proof of universality, Johansson observed that the matrices~$H_t$, $t\ge 0$, could be thought of in terms of Dyson's Brownian motion (DBM) model, where $H_t$ is obtained from $H$ by letting the matrix elements execute independent Brownian motions, subject only to Hermitian symmetry, for a~time~$t$. While Johansson did not utilize this observation directly in his proof, it turned out that the observation played, and continues to play, a central role, in the analysis of universality phenomena in random matrix theory. The realization that Dyson Brownian motion was related to the issue of universality can already be traced to the 1962 paper~\cite{Dyson62} where Dyson introduced DBM. Instead of Brownian motion, Dyson considered the process $\tH_t$ obtained by letting the entries of~$\tH_t$ evolve as independent Ornstein--Uhlenbeck processes. The process $\tH_t$ has an invariant measure, and most fortunately, that measure is precisely GUE. As $t\to \infty$, or, as it turns out, for f\/ixed $t$ as $n\to\infty$, the distribution for $\tH_t$ converges to the f\/ixed point, equilibrium distribution, viz., GUE! Putting aside the dif\/ferences between the Brownian process and the Ornstein--Uhlenbeck process, Dyson's observations provided compelling motivation for Johansson's result. The breakthrough came in 2009 when two independent groups, L.~Erd\H{o}s, S.~Peche, J.~Ramirez, B.~Schlein and H.-T.~Yau~\cite{EPRSY}, on the one hand, and T.~Tao and V.~Vu~\cite{TaoVu} on the other, almost simultaneously announced proofs of universality in the bulk for broad classes of Hermitian Wigner ensembles.

The basic idea in Yau et al.\ is to use DBM $\tH_t$ as above. As in the non-Ornstein--Uhlenbeck case considered by Johansson, for f\/ixed $t>0$, the distribution of the eigenvalues of $\tH_t$ in the bulk exhibit universal behavior as $n\to \infty$. However, as emphasized by Dyson in his 1962 paper, the time $t \sim 1$ is needed to reach global equilibrium for \textit{all} the eigenvalues of $\tH_t$: In order to reach local equilibrium, times $t \sim \frac{1}{n}$ should be suf\/f\/icient. On the other hand, for such small times $t\sim \frac{1}{n}$, or a little longer, one expects the eigenvalue distribution of $\tH_t$ to be essentially the same as the distribution of the eigenvalues of $H$. So one seeks a window given by $0<b<a<1$ such that for times $\le t^{-b}$, the eigenvalue distribution of $\tH_t$ is essentially the same as that of~$H$, while for times $\ge t^{-a}$, the eigenvalues of $\tH_t$ are already in local GUE equilibrium. As $\frac{1}{t^a} < \frac{1}{t^b}$, one draws the conclusion that the eigenvalues of $H$ have the same local distribution as GUE and it is this window that Yau et al.\ exploited to prove universality in the bulk for Hermitian Wigner matrices. Over the years this technique, now called the ``3 step'' method, has been greatly extended and ref\/ined by Yau and his many collaborators, and has now been used to prove universality in the bulk, and also at the edge, for orthogonal and symplectic ensembles, and more generally for certain $\beta$-ensembles for any $\beta>1$, certain Erd\H{o}s--Renyi random graphs and certain banded random matrices, and also, most recently \cite{AEK} for certain random matrices with short range correlations.

The basic idea of Tao--Vu to prove universality for Hermitian Wigner ensembles was to use Lindeberg's replacement trick. In 1922, Lindeberg gave a new proof of the central limit theorem for sums $S_n = x_1 + \dots + x_n$ of iid variables $\{x_i\}$ with distribution $\chi$, by replacing the $x_i$'s one at a time by iid Gaussian variables $\{\hx_i\}$ with the same mean and variance as $\chi$,
\begin{gather*}
S_n = x_1 + x_2 + \dots + x_n \to \hx_1 + x_2 + \dots + x_n \to \hx_1 + \hx_2 + x_3 + \dots + x_n\\
\hphantom{S_n}{}\to \dots \to S_n = \hx_1 + \hx_2 + \dots + \hx_n
\end{gather*}
with negligible error as $n\to \infty$. By analogy, starting with a Hermitian Wigner matrix $H=\{H_{ij}\}$, Tao and Vu showed that the $H_{ij}$'s with distribution $\chi$ could be replaced one at a time by iid Gaussian variables with the same mean and variance as~$\chi$, obtaining in the end, with negligible error in the bulk eigenvalue statistics as $n\to\infty$, a GUE matrix~$\hH$.

An open problem of great, ongoing interest is the question of the behavior of random $n\times n$ band matrices $H$ with bandwidth $w \le n$, $H_{ij} =0$ if $|i-j|\ge w$. The long-standing \textit{conjecture} of V.~Fyodorov and A.~Mirlin \cite{FyMi} from 1991 is that for iid $H_{ij}$, $ |i-j|< w$, with $w \gg \sqrt{n}$, the local statistics of the eigenvalues of $H$ are GUE, but for $w \ll \sqrt{n}$, the behavior is Poisson. Indeed for $w= n$, $H$ is a GUE matrix, and for $w=2$, the tridiagonal (Jacobi) case, results proving Poisson statistics go back to the 90's (see \cite{Minami}). This conjecture is of interest both in the real symmetric case, $H=\oH= H^T$, and in the complex Hermitian case, $H=H^\ast$. The primary obstacle in addressing the problem is that the powerful methods of invariant ensembles and the ``3-step''method of Yau et al.\ using DBM, as well as the replacement method of Tao--Vu, are no longer applicable/ef\/fective. New methods are needed. The only rigorous result indicating the critical~$\sqrt{n}$ threshold is due to T.~Shcherbina~\cite{TSh1,TSh2} concerning mixed moments of the characteristic polynomial for certain Gaussian ensembles. Shcherbina used so-called super-symmetric methods f\/irst introduced in the physics community, and then pioneered in the mathematical community by T.~Spencer, and which are, unfortunately, applicable only to Gaussian ensembles.

\looseness=-1 In all approaches to proving universality in the bulk for general Wigner ensembles $\{H\}$, the behavior of the eigenfunctions of the matrices $H$ plays an important role: Localized eigenfunctions are associated with Poisson statistics and delocalized eigenfunctions are associated with random matrix statistics. For banded random matrices, localization of the eigenfunctions has been proved for $w \ll n^{\frac{1}{8}}$~\cite{Schenk} and delocalization of the eigenfunctions for $w \gg n^{4/5}$~\cite{EKYY}. Recently~\cite{BEYY} have found a way to overcome some of the obstacles to implementing the ``3-step'' process, and they have proved bulk universality for random band matrices with independent entries in the case $w=\frac{1}{p} n$ for some f\/ixed integer~$p$. This is where the matter stands as of 2016 regarding bulk universality . On the other hand, Tracy--Widom statistics have been proved for extreme eigenvalues near the edge of the spectrum for $w \gg n^{5/6}$, an essentially optimal result~\cite{Sodin}.

Perhaps the most challenging problem currently in the spectral theory of Schr\"{o}dinger operators $H=-\De +V$, $\De= \text{Laplacian}$, is to analyze the spectrum of $H$ when $V$ is a random function. On the lattice $\ZZ^d$, a standard model (the Anderson model) is to assume that $\{V(x)\}_{x\in \ZZ^d}$ are iid. In seminal work in 1983, J.~Fr\"{o}hlich and T.~Spencer \cite{FrSp} showed that near the edge of the spectrum of~$H$, the eigenfunctions are localized. It is believed that in the bulk of the spectrum for $d=3$, the eigenfunctions of~$H$ are delocalized and the eigenvalues exhibit random matrix statistics. There is also believed to be a so-called mobility edge $\mu$ in the spectrum at which the statistics switch from a Poisson distribution to a random matrix distribution. So far this \textit{conjecture} is wide open. It is of interest to construct and analyze models where such a transition, Poisson $\to$ random matrix statistics, takes place, and apart from the intrinsic interest in random band matrices, much of the interest in random band matrices and the putative $\sqrt{n}$ threshold, lies in the role of random band matrices as models for localization/delocalization in random Schr\"{o}dinger theory. An interesting model that exhibits the Poisson-GUE transition was introduced by O.~Bohigas and M.~Pato~\cite{BoPa} who considered the gap probability of a random particle system obtained from the Coulomb gas by dropping at random a fraction $1-\ga$ of levels, a procedure known as ``thinning''. As $\ga \downarrow 0$ the gap probability becomes Poissonian, and as $\ga\uparrow 1$ the gap probability follows GUE (see also~\cite{BDIK} for a rigorous analysis). More information on the thinning process in RMT can be found, in particular, in~\cite{ChCl}.

Another \textit{open problem} in RMT is to prove universality, in the bulk and at (all) the edges, for Wigner ensembles in the case that the equilibrium measure for the ensemble is supported on more than one interval. Furthermore, one would like to analyze and control situations where parameters in the underlying matrix distributions vary, and intervals in the support of the equilibrium distribution for the ensemble open and close. By contrast with the Wigner case, these problems are well understood for unitary invariant ensembles, and to a slightly lesser extent, for invariant orthogonal and symplectic ensembles. For example, for the unitary invariant ensemble with probability distribution proportional to $e^{-n \operatorname{tr} V(M)} dM$, where $V(x)=-ax^2+bx^4$, with $b>0$ f\/ixed and $a\in \RR$, there is a critical value~$a_c$ such that for $a< a_c$, the equilibrium measure for the ensemble is supported on one interval, and for $a>a_c$ the measure is supported on two intervals. In both cases, the eigenvalues exhibit universal behavior in the bulk away from the end points of support of the equilibrium measure. In the critical case $a=a_c$, when the two intervals merge at some point $x=x_c$, the behavior of the eigenvalues in a~neighborhood of $x_c$ corresponds to a dif\/ferent universality class, which can be analyzed and described explicitly using, for example, the results in~\cite{DKMVZ}. Similar problems are completely \textit{open} for Wigner ensembles.

\looseness=-1 Double-scaling limits and, more generally, multi-scaling limits, are of great interest in physics and in mathematics. For example, regarding the classical Ising model in two dimensions, much work has been done on analyzing the double-scaling limit for the two-point correlation function as the temperature $T$ goes to the critical temperature $T_c$ and simultaneously the separation $x$ between the spins goes to inf\/inity. It turns out, quite remarkably, that the limit is described by a particular solution of the Painlev\'{e}~III equation~\cite{WMTB}. Apart from the intrinsic interest in this result, it is widely believed in the physics community that the behavior of the correlation function in the double-scaling limit is universal, and precisely the same behavior should be expected for a~very wide class of statistical spin models. In other words, double-scaling limits can provide access to universal behavior that transcends the problem at hand. Indeed, the random band mat\-rix problem described above as the band width $w$ and the mat\-rix size $n$ go to inf\/inity, provides a~model from which we hope to learn about localization in the bulk for random Schr\"{o}dinger ope\-ra\-tors. In RMT, various double-scaling problems have been considered. The f\/irst rigorous analysis is due to P.~Bleher and A.~Its \cite{BlIt} who considered the unitary invariant ensemble with potential $V(x) =-ax^2 +bx^4$ as above, in the limit as $a$ converges to the critical value $a_c$ and the size $n$ of the matrices goes to inf\/inity. Again Painlev\'e equations and associated functions arise in the analysis. In 2006, T.~Claeys and A.~Kuijlaars \cite{ClKu} generalized the results of Its and Bleher to a wider class of potentials $V(x)$, and in the present volume T.~Claeys and B.~Fahs \cite{ClFa} have analyzed a very interesting double-scaling problem where the Painlev\'e~I equation plays a role. There have also been numerous non-rigorous analyses of double-scaling problems in RMT, both within the mathematical community and the physics community. In RMT almost all problems involve matrix distributions which depend on some external parameters, say~$p$. Typically there are critical values $p_c$ at which the nature of the equilibrium measure for the ensemble changes. The double-scaling problems, $p\to p_c$ and matrix size $n\to \infty$, are of basic interest. Most of these problems remain \textit{open}.

An \textit{open} problem of long-standing interest concerns explicit formulae for the Tracy--Widom distribution functions for general $\beta$-ensembles. By the universality results mentioned above, we know that for each $\beta$, the distribution is universal, but up till now explicit expressions were known only for the invariant ensembles, $\beta=1, 2$ and $4$. However, recently Rumanov \cite{Rum2, Rum1} derived an explicit formula for the Tracy--Widom distribution in the case $\beta=6$ in terms of Painlev\'e~II and the solution of an auxiliary ODE. Although parts of Rumanov's analysis are not rigorous (see~\cite{GIKM} where some of the gaps are f\/illed), Rumanov's work constitutes a very signif\/icant development. As noted by Rumanov, his analysis extends, in principle, to the Tracy--Widom distributions for all even $\beta>0$. It is of great interest to \textit{complete} Rumanov's analysis rigorously, not only for $\beta=6$, but for all even $\beta>0$. Using loop equation methods, G.~Borot, B.~Eynard, S.~Majumdar and C.~Nadal~\cite{BEMN} have written down very detailed formulae for the asymptotic behavior of the Tracy--Widom distributions for general $\beta>0$. It is a challenging \textit{open} problem to verify rigorously these formulae of Borot et al.

There are two aspects to random matrix theory. The f\/irst aspect concerns internal questions within RMT, and most of the problems that we have considered so far, are of this nature. The second aspect concerns applications of RMT. The situation is analogous to the theory of orthogonal polynomials, where researchers study orthogonal polynomials per se, on the one hand, and their applications, on the other. As new applications arise, they can, and often do, raise new structural questions within the theory of orthogonal polynomials. The same is true for RMT, with RMT playing the role of a ``stochastic special function theory'', so extending the reach of classical special function theory to the nonlinear problems of modern mathematical physics. In Problem~\ref{prob3} we will discuss applications of RMT to interacting particle systems and in Problem~\ref{prob4} we will discuss applications of RMT to numerical analysis. As we will see, numerical analysis raises, in particular, many new, deep and very interesting problems in RMT.

\begin{problem}[interacting particle systems and KPZ]\label{prob3}
In 1999, J.~Baik, P.~Deift and K.~Johansson~\cite{BDJ} solved the longest increasing subsequence problem, establishing en route a connection between combinatorics and RMT (Tracy--Widom distribution). Shortly thereafter, many statistical models/interacting particle systems were solved in related ways, f\/irst by K.~Johansson~\cite{JohShape}, and then by A.~Borodin, A.~Okounkov and G.~Olshanski \cite{BOO}, among many others. All these early examples were ``integrable'' in the sense that certain powerful algebraic tools, e.g., determinantal particle systems, the Robinson--Schensted--Knuth correspondence, Fredholm operator theory, $\dots$, could be applied to reduce the problems at hand to analytically viable forms from which their asymptotic behavior could be deduced using standard asymptotic methods such as the Laplace method/stationary phase, or sometimes, Riemann--Hilbert/nonlinear steepest-descent methods. In a~very signif\/icant and unanticipated development during 2007--2009, C.~Tracy and H.~Widom \cite{TW2,TW1,TW3} analyzed ASEP (the asymmetric simple exclusion process) with step initial data, a system to which the above algebraic tools cannot be applied. Nevertheless, in an algebraic/combinatorial/analytical tour de force, they were able to show that ASEP exhibited the same asymptotic RMT behavior as the integrable models.

A striking feature of all these models, is that f\/luctuations are of order $N^{1/3}$, where $N\to \infty$ is the size of the system, thereby showing that the f\/luctuations for these models are in the KPZ (Kardar--Parisi--Zhang) universality class. Note that, by contrast, systems in the Gaussian universality class have f\/luctuations of order $N^{1/2}$. The notion of KPZ universality comes from the work of M.~Kardar, G.~Parisi and Y.-C.~Zhang~\cite{KPZ} who in 1986 introduced an equation, the KPZ equation, to describe the growth of random interfaces with height function $h(x,t)$, $x\in \RR$, $t\ge 0$. The KPZ equation is stochastic and non-linear, and using various physical arguments they showed that as $t \to \infty$, the surface should be correlated over distances of order $t^{2/3}$ and the f\/luctuations of $h$ should be of order~$t^{1/3}$. In other words there should be a $3:2:1$ scaling for time~: spatial correlations~: f\/luctuations. It is widely believed that a large class of physical systems and mathematical models obey KPZ $3:2:1$ universality. The above integrable models were the f\/irst to be shown rigorously to have KPZ f\/luctuations $3:1$ (i.e., $N:N^{1/3}$). In addition, in 2000, K.~Johansson \cite{JohanssonTrans} showed that for the last passage percolation problem (LPP), spatial correlations were of order $t^{2/3}$ in accord with $3:2$ KPZ universality, and in later more detailed work Johansson \cite{JohAiry} showed that as $t\to \infty$ the f\/ixed $t$ marginal for $h(x,t)$ converged to the Airy~2 process. We note that M.~Pr\"ahofer and H.~Spohn \cite{PS} were the f\/irst to show that $h(x,t)$ converged to the Airy~2 process, but their argument was not rigorous (see~\cite{ICreview} for more information on spatial correlations).

\looseness=1 A basic question is whether KPZ is in the KPZ universality class. In other words, can one show rigorously that solutions of KPZ indeed have the KPZ scaling $3:2:1$? The f\/irst dif\/f\/iculty here is to show that the Cauchy problem for the non-linear stochastic KPZ equation indeed has a solution. Following L.~Bertini and N.~Cancrini \cite{BC}, one approach is to use the Cole--Hopf transformation which converts the KPZ equation into a heat equation with a random potential which can then be solved via a generalized Feynmann--Kac formula. In breakthrough work in 2010, G.~Amir, I.~Corwin and J.~Quastel~\cite{ACQ}, used results from the 2008--2009 work of Tracy and Widom \cite{TW2,TW1,TW3} on ASEP to f\/ind an explicit formula for the solution of the Cole--Hopf transformed KPZ equation with so-called ``narrow wedge'' initial data. From this formula they were able to extract the $t^{1/3}$ KPZ scaling for the f\/luctuations of the solution, but their methods give no information on spatial correlations. Simultaneous with the work of Amir, Corwin and Quastel on the KPZ equation, there was also the independent, related, but not rigorous work, of T.~Sasamoto and H.~Spohn~\cite{SaSp} also relying on Tracy--Widom's ASEP formula, as well as the work of V.~Dotsenko \cite{Dot}, and P.~Calabrese, P.~Le~Doussal and A.~Russo~\cite{CDR} using replicas. The varying levels of rigor of these papers are discussed in~\cite{ICreview}. Regarding full $3:2:1$ KPZ universality, the strongest result is due to I.~Corwin and A.~Hammond \cite{CH2} (see~\cite{CQR} for the strongest and most precise statement of the KPZ universality conjecture for the KPZ equation). Again for the stochastic heat equation with narrow wedge initial data, Corwin and Hammond were able to show tightness of the suitably $3:2:1$ scaled height function $h(x,t)$ in a ``good'' sense which implied convergence for suitable subsequences to limiting processes with ``good'' properties, but they were not able to show that the limits were unique and given by the physically anticipated process. In particular, they could not identify the anticipated Airy$_2$ marginal. We note here that M.~Hairer \cite{Hairer} was able to show directly without using the Cole--Hopf transformation that the Cauchy problem for the KPZ equation has a~solution. This was a~major achievement that f\/igured prominently in the citation for Hairer when he won the Fields Medal in 2014. Under the appropriate conditions, Hairer's solution agrees with a Cole--Hopf solution. For more information on KPZ, see, for example \mbox{\cite{BorCor, BorPet, ICreview, ICAMS}}.

The most challenging \textit{open} problem regarding the KPZ equation is to prove $3:2:1$ KPZ universality in the strong sense of Corwin, Quastel and Remenik~\cite{CQR} alluded to above, for solutions of the Cauchy problem of the KPZ equation for a wide set of initial data, and to identify the limiting process. Another \textit{open} problem is to analyze solutions of KPZ at more than one space-time point. Regarding these issues, much signif\/icant progress has been made for various integrable particle systems. For example in 2015 K.~Johansson \cite{JohTwo} computed correlations for two space-time points for the Brownian LPP problem, from which it easily follows, in particular, that Brownian LPP exhibits $3:2:1$ KPZ scaling. We note that this is the only system for which $3:2:1$ KPZ scaling has been verif\/ied rigorously. Also, in 2015, I.~Corwin, Z.~Liu and D.~Wang \cite{CLW} demonstrated KPZ universality for the f\/luctuations for LPP and TASEP with general initial data. However, many basic problems remain \textit{open}. In particular showing that LPP with general weights has $3:2:1$ scaling, is far from being resolved. Also, regarding the original longest increasing subsequence problem, KPZ universality for the f\/luctuations has been proved so far only for the uniform distribution on the permutations: The problem for the f\/luctuations with more general distributions is very much \textit{open}. In another direction, it remains an \textit{open}, and somewhat mysterious, problem to understand how Tracy and Widom's solution of ASEP actually ``works'', and to extend their method of solution to more general initial data (however, see \cite{BorCor}, and more recently,~\cite{BorOlshASEP}).

Finally we note that, with few exceptions, all the rigorous asymptotic scaling results that have been obtained so far for KPZ/interacting particle systems, have been obtained only for restricted classes of initial data. In addition, it is particularly frustrating that there is no known perturbation theory, analogous to KAM theory for integrable dynamical systems, that one can use to investigate the persistence of ``integrability'' for non-solvable models and demonstrate robustness for asymptotic scaling phenomena. Developing such a perturbation theory is an \textit{open} conceptual and technical problem of the f\/irst order.
\end{problem}

\begin{problem}[numerical computation with random data]\label{prob4}
Standard algorithms to compute the eigenvalues of a random matrix $H$ are completely integrable Hamiltonian systems (see \cite{DLNT, DLT,Symes}). The situation is as follows. Let $\Sg_n$ denote the set of real $n\times n$ symmetric matrices. Associated with each algorithm~$\mA$, there is, in the discrete case such as the QR algorithm, a~map $\varphi = \varphi_{\mA}\colon \Sg_n \to \Sg_n$, with the properties
\begin{alignat*}{3}
&\bullet \text{(isospectral)} \quad && \text{spec }(\varphi_{\mA}(H)) = \text{spec $(H)$}, & \\
&\bullet \text{(convergence)} \quad && \text{the iterates} \ X_{k+1} = \varphi_{\mA} (X_k), \ k\ge 0, & \\
&&& X_0 = H \quad \text{given, converge to a diagonal matrix $X_\infty$ as $k\to\infty$},&
\end{alignat*}
and in the continuum case, such as the Toda algorithm, there is a f\/low $t\to X(t) \in \Sg_n$ with the properties
\begin{alignat*}{3}
&\bullet \text{(isospectral) } \quad && \text{spec}(X(t)) \text{ is constant, $t\ge 0$},& \\
&\bullet \text{(convergence)} \quad && \text{the f\/low $X(t)$, $t\ge 0$, $X(0)=H$ given},& \\
& && \text{converges to a diagonal matrix $ X_\infty, X(t)\to X_\infty$ as $ t\to\infty$}.&&
\end{alignat*}
In both cases, necessarily the (diagonal) entries of $X_\infty$ are the eigenvalues of the given matrix~$H$. The f\/low $X(t)$ in the continuum case is completely integrable and in the discrete case $X_k$ is the $\text{time} - k$ evaluation of a completely integrable f\/low (see references above). Given $\ep >0$, it follows, in the discrete case, that for some $m$ the of\/f-diagonal entries of $X_m$ are $O(\ep)$ and hence the diagonal entries of $X_m$ give the eigenvalues of $X_0=H$ to $O(\ep)$. The situation is similar for continuous algorithms $t\to X(t)$.

On the other hand, matrices carry another natural integrable structure. This is the structure of RMT in which basic statistics, such as the gap probability or the nearest-neighbor distribution, converge as the size $n$ of the matrices becomes large to distributions that are expressed in terms of the solution of integrable Hamiltonian systems (in particular, Painlev\'e equations). The question raised in \cite{PercyCM} was: What happens if we try to marry these two integrabilities? Or more precisely, what can be said about the computation of the eigenvalues of random matrices using standard algorithms?

This question was taken up by P.~Deift, G.~Menon and C.~Pfrang in 2012 \cite{DMP}. What they found was that the computations exhibited universality, with a dif\/ferent universality class for each algorithm $A$. More precisely, for an $n\times n$ matrix $H$ chosen from an ensemble $\mE$ on $\Sg_n$, and a given accuracy $\ep >0$, let $T(H)= T_{\ep,n,\mA,\mE}(H)$ be the smallest $k$, or smallest $t$ in the continuum case, for which $X_k$ with $X_0=H$, evolving under algorithm $\mA$, has of\/f diagonal entries (see ``def\/lation time'' in~\cite{DMP} and also $T^{(1)}$ below) less than $\ep$. The authors in~\cite{DMP} then recorded
\begin{gather*}
\tau_{\ep, n, \mA, \mE}(H) \equiv \frac{T_{\ep, n, \mA, \mE}(H)- \llg T_{\ep, n, \mA, \mE}\rrg}{\sg_{\ep, n, \mA, \mE}},
\end{gather*}
where $\llg T_{\ep, n \mA, \mE}\rrg$ and $\sg^2_{\ep, n, \mA, \mE}$ are respectively the sample average and sample variance for \linebreak $T_{\ep, n,\mA, \mE}(H)$ computed for a~very large sample of matrices $H\in \mE$. What the authors found was that, for a~given algorithm~$\mA$, and~$\ep$,~$n$ in a suitable scaling region, the histogram for $\tau_{\ep, n, \mA, \mE}$ was \textit{universal}, independent of the ensemble~$\mE$. This phenomenon was present, in particular, for such disparate distributions as GOE and Bernoulli. The histogram for $\tau_{\ep, n, \mA, \mE}$ does, however, depend on~$\mA$ and is dif\/ferent, in particular, for the~$QR$ and Toda algorithms.

In \cite{DMOT} the authors P.~Deift, G.~Menon, S.~Olver and T.~Trogdon raised the question of whether the universality results in~\cite{DMP} were limited to eigenvalue algorithms, or whether they were present more generally in numerical computations. And indeed in~\cite{DMOT} the authors found similar universality results for a wide variety of numerical algorithms, including
\begin{itemize}\itemsep=0pt
\item more general eigenvalue algorithms such as the Jacobi algorithm, and also algorithms for Hermitian ensembles,
\item the conjugate gradient and GMRES algorithms to solve linear $n\times n$ systems $Hx=b$, $H$~and~$b$ random,
\item an iterative algorithm to solve the Dirichlet problem $\De u=0$ on a random star-shaped region $\Om \subset \RR^2$ with random boundary data $f$ on $\p \Om$, and
\item a genetic algorithm to compute the equilibrium measure for orthogonal polynomials on the line, and was also for a decision making process in recent laboratory experiments of Yu.~Bakhtin and J.~Correll \cite{BakCor}.
\end{itemize}

Until recently, all the evidence for universality in computation was numerical and experimental. In order to establish the phenomenon rigorously, P.~Deift and T.~Trogdon \cite{DT} focused on the Toda algorithm, $\mA=\text{Toda}$, applied to $n\times n$ Hermitian matrices $H$ chosen from some ensemble $\mE$, with stopping time $T^{(1)}(H)=T^{(1)}_{\ep, n, \text{Toda}, \mE}(H)$, which is the f\/irst time that
\begin{gather*}
\sum^n_{j=2} |X_{1j}(t)|^2 = \sum^n_{i=2} |X_{i1}(t)|^2 < \ep^2,
\end{gather*}
where $X(t)$ solves the Toda f\/low with $X(0)\!=\!H$. For $t\!=\!T^{(1)}(H)$ we clearly have \mbox{$|X_{11} (t)\!-\!\lambda^\ast|\!< \!\ep$} for some eigenvalue $\la^\ast$ of $H$. By the ordering properties of the Toda algorithm, $\la^\ast=\la_{\max}$, the largest eigenvalue of $H$, with high probability as $n\to\infty$. Thus the Toda algorithm with stopping time $T^{(1)}(H)$ is an algorithm to compute the largest eigenvalue of a given matrix~$H$.
\end{problem}

In \cite{DT} the authors proved the following result:
\begin{Theorem}[universality for the Toda algorithm]\label{thm:1}
Let $0<\tau <1$ be fixed and let $(\ep, n)$ be in the scaling region
\begin{gather*}
\frac{\log \ep^{-1}}{\log n} \ge \frac{5}{3} + \frac{\tau}{2}.
\end{gather*}
Then if $H$ is distributed according to any real $(\beta=1)$ or complex $(\beta=2)$ invariant or Wigner ensemble\footnote{For a precise description of the ensembles allowed, see \cite{DT}.}
\begin{gather*}
\lim_{n\to\infty} \prob \left(\frac{T^{(1)}}{C^{2/3}_v 2^{-2/3} N^{2/3}
\left(\log \ep^{-1}-\frac{2}{3} \log n\right)} \right) = F^{\text{\rm gap}}_{\beta} (t).
\end{gather*}
\end{Theorem}

Here $C_v$ is an explicit ensemble dependent constant and for $\beta=1,2$,
\begin{gather*}
F^{\text{gap}}_{\beta}(t) = \lim_{n\to\infty} \prob \left(\frac{1}{C^{2/3}_v 2^{-2/3} n^{2/3}(\la_n-\la_{n-1})} \le t\right),
\end{gather*}
where $\la_{\max} = \la_n > \la_{n-1} > \dots > \la_j > \dots > \la_1$ are the eigenvalue of a random matrix~$H$. Properties of $G_\beta(t) \equiv 1-F_\beta^{\text{gap}} (t)$, the distribution function for the f\/irst gap are examined, for example, in~\cite{WBF}. The authors also prove as a corollary that $\ep^{-1} \big|\la_{\max} - X_{11} \big(T^{(1)}\big) \big|$ converges to zero as $n\to\infty$ in the scaling region.

The Toda system has the form
\begin{gather*}
\frac{dX}{dt}= [X, B(X) ], \qquad B(X)= X_- - X^*_-, \qquad X(0)=H,
\end{gather*}
where $X_-$ is the strictly lower-triangular part of $X$, $X^*_-$ is the adjoint of $X_-$, and $[A,B]$ is the standard matrix commutator. The proof of the above theorem relies in a crucial way on the integrability of the Toda system which makes it possible to reduce the proof to the analysis of explicit formulae. Indeed, following Moser \cite{Moser} one f\/inds that
\begin{gather}
E(t) \equiv \sum^n_{j=2} |X_{ij}(t) |^2 = \sum^n_{j=1} (\la_j - X_{11}(t) )^2 |u_{1j}(t) |^2,\label{eq4}\\
\la_j - X_{11}(t) = \sum^{n-1}_{i=1} (\la_j-\la_i ) |u_{1i} (t) |^2, \label{eq5}
\end{gather}
where $u_j(t) = (u_{1j} (t), \dots, u_{nj}(t) )^T$ is the $j^{\rm th}$ normalized eigenvector of $X(t)$, $ X(t) u_j(t)=\la_j u_j(t)$, which evolves as
\begin{gather}\label{eq6}
|u_{1j}(t) | = \frac{\beta_j e^{\la_j t}}{\left(\sum\limits^n_{i=1} \beta^2_i e^{2\la_i t}\right)^{\frac{1}{2}}},
\end{gather}
where $\beta_j= |u_{ij}(0) |$ are the moduli of the f\/irst components of the eigenvectors of $X(0)=H$. Substituting \eqref{eq5}, \eqref{eq6} into \eqref{eq4} we obtain an explicit formula for~$E(t)$ in terms of the eigenva\-lues~$\{\la_j\}$ and the moduli $\{\beta_j\}$ of the f\/irst components of the normalized eigenvectors of~\mbox{$H=X(0)$}. To obtain $T^{(1)}$, we must solve for the f\/irst $t$ such that $E(t) = \ep^2$. We see, in particular, that the statistics of $T^{(1)}$ depends on the statistics of $\{\la_j\}$ and $\{\beta_k\}$. More precisely, in order to analyze the statistics of $T^{(1)}$, it turns out that one needs to control the probabilities for the following events as $n\to\infty$:
\begin{condition}\label{con1}
For $0<p< \frac{1}{3}$
\begin{enumerate}\itemsep=0pt
\item[(i)]$\la_{n-1}- \la_{n-2} \ge p (\la_n - \la_{n-1} )$.
\end{enumerate}
Let $G_{n,p}$ denote the set of matrices that satisfy this condition.
\end{condition}
\begin{condition}\label{con2}
For any f\/ixed $0<s< \min \{\sg/44, p/8\}$
\begin{enumerate}\itemsep=0pt
\item[(i)] $\beta_j \le n^{-\frac{1}{2} + s/2}$ for all $j$,
\item[(ii)] $n^{-\half - s/2} \le \beta_j$ for $j = n, n-1$,
\item[(iii)] $n^{-2/3 - s} \le \la_n-\la_j \le n^{-2/3 + s }$ for $j=n-1, n-2$, and
\item[(iv)] $ | \la_j - \ga_j | \le n^{-2/3 + s} ( \min \{ j, n-j + 1\} )^{-1/3}$ for all $j$.
\end{enumerate}
Here the $\ga_j$'s represent the quantiles of the equilibrium measure $d\mu$, i.e., $\ga_j$ is def\/ined to be the smallest value of $b$ such that $j/n=\int^b_{-\infty} d\mu$. Let $R_{n,s}$ denote the set of matrices that satisfy these conditions (i)--(iv).
\end{condition}

In order to prove Theorem~\ref{thm:1} and the corollary, one needs to show that for both Wigner ensembles (WE's) and invariant ensembles (IE's), Condition~\ref{con1} holds with probability~1 as $p\downarrow 0$, i.e.,
\begin{gather}\label{eq7}
\lim_{p\downarrow 0} \limsup_{n\to\infty} \prob \big(G^{ c}_{n p}\big) =0
\end{gather}
and Condition~\ref{con2} holds with high probability as $n\to\infty$, that is, for any $s>0$
\begin{gather}\label{eq8}
\prob (R_{n, s} )=1+o(1)
\end{gather}
as $n\to\infty$.

In addition one needs the following results: For both WE's and IE's, as $n\to\infty$
\begin{gather}\label{eq9}
n^{-1/2} \{\beta_n, \beta_{n-1}, \beta_{n-2} \}
\end{gather}
converge jointly in distribution to $ \{ |X_1|, |X_2|, |X_3| \}$ where $\{X_1, X_2, X_3\}$ are iid real $(\beta=1)$ or complex $(\beta=2)$ standard normal random variables. Additionally
\begin{gather}\label{eq10}
2^{-2/3} n^{2/3} ( b_v- \la_n, b_v-\la_{n-1}, b_v-\la_{n-2} )
\end{gather}
converge jointly in distribution to random variables $ \{\La_{1,\beta}, \La_{2,\beta}, \La_{3,\beta} \}$ which are the smallest three eigenvalues of the stochastic Airy operator. Furthermore $ (\La_{1,\beta}, \La_{2,\beta}, \La_{3,\beta} )$ are distinct with probability one. Here $b_v$ is the top of the support of the equilibrium measure $d\mu$. In order to prove \eqref{eq7}--\eqref{eq10}, one needs to draw on results from the forefront of research in RMT and that were proved only very recently. In particular, one needs to draw on results from \cite{BEY,BY, Erdos, EYY, RRV}. How these results are used to prove \eqref{eq7}--\eqref{eq10} is described in detail in~\cite{DT}.

It is clear from the above considerations that the analysis of numerical algorithms with random data is a rich source of new and challenging problems in RMT with immediate practical applications. Here are some \textit{open} problems. In order to prove universality for the centered and scaled variable $\frac{T^{(1)}- \llg T^{(1)} \rrg}{\sg_{T^{(1)}}}$ (cf.\ $\tau_{\ep,n,\mA,\mE}$ at the beginning of this section) one needs to control the statistics of the $\la_j$'s and the $\beta_k$'s as in \eqref{eq9}, \eqref{eq10} \textit{jointly} as $n\to\infty$. For $\beta=2$ invariant ensembles, this can be done, but for $\beta=1$ invariant ensembles, and for $\beta=1$ or 2 WE's, this problem is still very much \textit{open}. For the $QR$ algorithm, the role of the eigenvalue gaps $\la_k-\la_j$ is now played by the eigenvalue ratios $\la_k/\la_j$. So instead of the behavior of $T^{(1)}=T^{(1)}_{\ep,n,QR,\mE}$ being controlled by the inverse gap $ (\la_n-\la_{n-1} )^{-1}$ as in Theorem~\ref{thm:1}, $T^{(1)}$ will now be controlled by $\big|\frac{\la_{j+1}}{\la_j}\big|$ and $\big| \frac{\la_{j-1}}{\la_j}\big|$ where $\la_j$ is the eigenvalue of $H$ closest to zero. It is very much an \textit{open} problem to control, as $n\to\infty$, the statistics of $\big|\frac{\la_{j+1}}{\la_j}\big|$, $\big| \frac{\la_{j-1}}{ \la_j} \big|$ for $\la_j \sim 0$ deep in the bulk of the spectrum, together with their associated normalized eigenvectors, as in \eqref{eq9},~\eqref{eq10}. The $QR$ algorithm is the ``go to'' algorithm for eigenvalue computation, and the proof of the analog of Theorem~\ref{thm:1} for the~$QR$ algorithm is an \textit{open} problem of great interest. For recent work in proving universality for the $QR$ algorithm and related algorithms for strictly positive def\/inite matrices, see~\cite{DT17}.

The most challenging problem for eigenvalue computation with random data is to prove the analog of Theorem~\ref{thm:1} for the def\/lation time $T_{\ep,n,\mA,\mE}$ (see \cite{DT} and the references therein) for some algorithm $\mA$. If $\mA= \text{ Toda}$, for example, there is an analog $E_{\tdef}(t)$ of $E(t)$ in \eqref{eq4}. Again one computes $T_{\ep,n,\mA,\mE}$ as the smallest time for which $E_{\tdef} (t) = \ep^2$, but now $E_{\tdef}(t)$ is expressed in terms of \textit{all} the components of the eigenvectors of $X(0)=H$, in addition to the eigenvalues. We refer the reader to \cite{DT} where one can see precisely what statistics on the eigenvalues and eigenvectors are needed to prove universality for $T_{\ep,n,\mA, \mE}$, and how the problem can be broken down into smaller sub-problems of interest. Indeed, $T_{\ep,n,\mA,\mE}$ is the smallest time $t$ for which $X(t)$ has the block form $\left(\begin{smallmatrix} A&B\\ B^* & C \end{smallmatrix} \right)$ where $A$ is $j\times j$, $C$ is $(n-j) \times (n-j)$ and $\|B\|= \|B^*\| \le \ep$ for some $1\le j \le n-1$. For such $t$, the eigenvalues of the \textit{deflated} matrix $\left(\begin{smallmatrix} A &0\\ 0&C \end{smallmatrix}\right)$ agree with the eigenvalues of $X(0)=H$ to $O(\ep)$. The algorithm with def\/lation then continues by applying $\mA$ to the (smaller) matrices $A$ and $C$, etc. Note that the time $T^{(1)}$ above controls def\/lation to the form $\left(\begin{smallmatrix} A & 0\\ 0 & C \end{smallmatrix}\right)$ where $A$ is $1-1$ and $C=(n-1) \times (n-1)$. The next step in analyzing $T_{\ep, n, \mA, \mE}$ is to control $T^{(2)}$, the time for def\/lation to the form $\left(\begin{smallmatrix} A&0\\ 0 &C \end{smallmatrix}\right)$ where $A$ is $2\times 2$ and $C\text{ is } (n-2) \times (n-2)$. To control $T^{(2)}$ would require further advances in RMT, in particular, statistical information on the second components of the eigenvectors, and so on.

It is an interesting and challenging open problem to prove universality rigorously for any of the other numerical algorithms discussed in~\cite{DMOT}. The conjugate gradient algorithm, in particular, to compute the solution of the linear system $Ax=b$ with $A$ and~$b$ random, seems to be ``within reach''. In this connection, we refer the reader to~\cite{DMT}.

Finally we consider an open statistical problem. In order to test a given algorithm for universality, one needs to sample various ensembles. This is a standard matter if one is dealing with Wigner ensembles and the entries are independent, but when the entries are dependent, as in the case for general invariant ensembles, this is a non-trivial matter. In 2013, X.~Li and G.~Menon \cite{LM} developed an approach for sampling invariant ensembles based on simulating Dyson Brownian motion. In 2014, for invariant unitary ensembles, S.~Olver, R.~Nadakuditi and T.~Trogdon \cite{ONT} developed an alternative approach based on computing the orthogonal polynomials naturally associated with the ensembles. While the method of Li and Menon is signif\/icantly faster, the method of Olver et al.\ samples the distribution to essentially machine accuracy, a task which is computationally impractical using Dyson Brownian motion. It is an \textit{open} problem to extend the method of S.~Olver et al.\ to orthogonal and symplectic invariant ensembles.

\begin{problem}[initial boundary value problems for integrable systems (IBVP)]\label{prob5}
Initial boundary value problems for integrable systems in $1+1$ dimensions are of great interest. In seminal work in the late 1990s A.~Fokas \cite{Fokas} introduced a new, more f\/lexible approach to inverse scattering, and over the years many researchers (A.~Fokas, A.~Its, A. Boutet de Monvel, D.~Shepelsky, \dots) have applied Fokas' approach to IBVPs for various integrable systems. It turns out, however, that in Fokas' approach the general initial value problem is overdetermined. For example, in order to solve NLS
\begin{gather*}
i u_t = u_{xx} \pm 2 |u|^2 u
\end{gather*}
in $(x\ge 0$, $t\ge 0)$ one needs to know $u_x(0,t)$, $t\ge 0$, in addition to the standard determining data
\begin{alignat*}{3}
& u(x,0) = f(x), \qquad && x \ge 0, &\\
& u(0,t) = g(t), \qquad && t \ge 0.&
\end{alignat*}
In certain ``integrable'' cases (see below) $u_x(0,t)$ can be determined a priori, but this is not the case in general. As demonstrated by Fokas, his approach leads to a Riemann--Hilbert reformulation of the IBVP, but one can only apply the powerful non-linear steepest descent method to determine the asymptotic behavior of $u(x,t)$ as $t\to \infty$ in the case that $u_x(0,t)$ is known a~priori. A~particular stand out \textit{open} problem is to compute the long-time behavior of the solution $u(x,t)$ of the focusing NLS equation, $i u_t= u_{xx} + 2|u|^2 u$, with $u(x,0)=f(x)$, $x\ge 0$, given and $u(0,t) = e^{iwt}$, $t\ge 0$ with $w\in \RR$. Interesting partial results for this problem as $w\to \infty$ are given in A.~Fokas and J.~Lenells \cite{FL}.

What kind of behavior do we anticipate for $u(x,t)$ as $t\to \infty$, $u(0,t) =e^{iwt}$? From a physical point of view one can think of a long rope which is shaken a one end with frequency $w/2\pi$. Experience tells us that for $w$ very large, only a very small disturbance will be generated in the rope. On the other hand, as we shake the rope slower and slower, more complicated and sizable wave forms propagate along the rope. In an analogous situation, P.~Deift, T.~Kriecherbauer and S.~Venakides \cite{DKV} considered a lattice of particles $x_k$, $ k\ge 0$, interacting with Toda forces
\begin{gather*}
\ddx_k = e^{x_{k-1}-x_k} - e^{x_k-x_{k+1}}, \qquad k\ge 1
\end{gather*}
and initial/boundary value conditions
\begin{gather*}
x_k(0) = \dx_k(0) = 0,\qquad k \ge 1,\\
x_0(t) = 2at + h(\ga t).
\end{gather*}
Here $h(\cdot)$ is $2\pi$-periodic and $a$ and $\ga>0$ are given constants, with $a$ below a certain critical value~$a_{\tcrit}$. The authors in~\cite{DKV} observed the following phenomena: In a frame moving with the average velocity of the drive $x_0(t)$, as $t\to\infty$ the asymptotic motion of the particles behind the shock front is $2\pi/\ga$-periodic in time,
\begin{gather*}
x_k \left(t + \frac{2\pi}{\ga}\right) = x_k(t), \qquad \text{for f\/ixed $k\ge 1$.}
\end{gather*}
Moreover, there is a sequence of thresholds
\begin{gather*}
\ga_1 > \ga_2 > \dots > \ga_j > \dots > 0, \qquad \ga_j \to 0 \quad\text{as} \quad j\to \infty
\end{gather*}
with the following properties:
\begin{itemize}\itemsep=0pt
\item If $\ga > \ga_1$, there exist constants $c, d$ such that $x_k-ck-d$ converges exponentially to zero for $1\ll k\ll t$. In other words, the ef\/fect of the oscillatory component of the driver does not propagate into the lattice and away from the boundary at $k=0$.
\item If $\ga_1 > \ga> \ga_2$, then the asymptotic motion is described by a travelling wave
\begin{gather*}x_k(t) = c_1 k + Y_1 (\beta_1 k + \ga t), \qquad 1 \ll k \ll t
\end{gather*}
transporting energy away from the driver. Here $c_1$ and $\beta_1$ are certain determined constants and $Y_1(1)$ is $2\pi$-periodic 1-gap solution of the periodic Toda lattice.
\item More generally, if $\ga_j > \ga > \ga_{j+1}$, a multi-phase wave emerges of the form
\begin{gather*}
x_k = c_j k + Y_j (\beta_1 k + \ga t, \beta_2 k + \ga t, \dots, \beta_j k + \ga t), \qquad 1 \ll k \ll t
\end{gather*}
again transporting energy away from the driver, for certain constants $c_j, \beta_1, \dots, \beta_j$. Now $Y_j(\cdot, \dots, \cdot)$ is $j$-gap solution of the periodic Toda lattice, $2\pi$-periodic in each of its variables.
\end{itemize}

Thus, at the phenomenological level, we see that indeed the periodically driven lattice behaves like a long rope which one shakes up and down at the one end. We expect similar behavior for focusing NLS, with $k$-gap solutions of periodic NLS emerging from the driver $e^{iwt}$ as $w\downarrow 0$. This is an \textit{open} problem of the f\/irst order, illustrating how the natural modes of an extended physical system ($k$-gap solutions in the case of NLS) can be excited by driving the system externally at one end.

In 2010, P.~Deift and J.~Park \cite{DP} considered the IPVP for focusing NLS with the boundary condition $u(0,t)=g(t)$ replaced by a Robin boundary condition,
\begin{gather*}
u_x(0,t) + q u(0,t)=0
\end{gather*}
with $q$ f\/ixed. This problem arises, in particular, in analyzing the Gross--Pitaevskii equation with a delta function external potential at $x=0$ and even initial data. It turns out that the NLS equation with Robin boundary conditions is an example of an integrable system which is not overdetermined from the point of view of Fokas' method, and indeed for this problem the method can be used to obtain a well-def\/ined and ef\/fective Riemann--Hilbert representation for the solution $u(x,t)$ of the IBVP (see~\cite{IS}). In their 2010 article, however, Deift and Park \cite{DP} used an earlier method introduced by R.~Bikbaev and V.~Tarasov in 1991~\cite{BT}. The method of Bikbaev and Tarasov is a non-linear version of the method of images in linear theory, where the solution in $x>0$ is extended to $x<0$, not by ref\/lection as in the linear case, but by a suitable B\"acklund transformation. In this way the IBVP is converted smoothly into a standard problem on the whole line, to which Riemann--Hilbert/steepest-descent methods apply directly. It is an interesting \textit{open} problem to use the Bikbaev--Tarasov method to solve the IBVP for defocusing NLS, $iu_t=u_{xx}-2|u|^2 u$, with Robin boundary conditions at $x=0$. In this case the B\"acklund transformation extending the solution from $x>0$ to $x<0$ introduces certain singularities which must be controlled and the long-time behavior of the solution of the IBVP will be very dif\/ferent from the case of focusing NLS. Another \textit{open} problem, which has broad implications, concerns the smoothness of the solutions $u(x,t)$ of the IBVP for NLS. One expects that if one solves the IBVP for NLS with $f(x)=u(x,0)$ and $g(t)=u(0,t)$ suf\/f\/iciently smooth, and with the right matching conditions at the corner $x=0$, $t=0$, then the solution $u(x,t)$ should be classical in $x>0$, $t>0$, in the sense that $u(x,t)$ is $C^2$ in $x$ and $C^1$ in $t$. However in the case of focusing NLS with smooth $f(x)=u(x,0)$ and Robin boundary conditions with $q\neq 0$ (the case $q=0$ is trivial~-- use simple ref\/lection), the only way in which Deift and Park are able to show that~$u(x,t)$ is a classical solution is by using the B\"acklund transform method of Bikbaev and Tarasov. It is easy to show that the solution has one spatial derivative in~$L^2$, but any attempt to iterate the solution in the usual way in some suitable Sobolev space, $\mH$ say, fails, as the mismatch between the non-linearity of the equation and the Robin boundary condition immediately kicks one out of the putative iterating space~$\mH$. This means, in particular, if one replaces the specif\/ic non-linearity $2|u|^2 u$ in the equation by any other reasonable and smooth non-linearity, for which there is no known B\"acklund method, for example $2|u|^4 u$, one does not know that the solution of the IBVP is classical. So the \textit{open} problem here is a PDE problem, viz., to prove that the IPVP for NLS with a general class of non-linearities, with Robin boundary conditions and matching, smooth initial data, is classical in $x>0$, $t>0$. Much more generally, the \textit{open} problem is to understand what can be said about smoothness of solutions of IBVP's for dispersive equations in any dimension. This problem seems to be largely unexplored.
\end{problem}

\begin{problem}[numerical solution of integrable systems]\label{prob6}
Solutions of the Cauchy problem for linear dispersive equations can be expressed in terms of Fourier integral operators. For example, the solution $u(x,t)$ of the free Schr\"{o}dinger problem
\begin{alignat*}{3}
& i u_t = u_{xx} , \qquad && -\infty < x < \infty,\quad t>0, &\\
& u(x,0) = f(x) , \qquad && -\infty< x< \infty &
\end{alignat*}
has the form
\begin{gather}\label{eq11}
u(x,t) = \frac{1}{\sqrt{2\pi}} \int \hf(z) e^{i (xz + tz^2)} dz,
\end{gather}
where $\hf(z) = \frac{1}{\sqrt{2\pi}} \int f(x) e^{-ixz} dz$ is the Fourier transform of $f$. As $t\to \infty$, we can eva\-luate~$u(x,t)$ using the classical steepest descent method, to obtain
\begin{gather}\label{eq12}
u(x,t) \sim e^{i\pi/4} \hf(z_0) \frac{e^{ix^2/4t}}{(2t)^{1/2}},
\end{gather}
where $z_0= -\frac{x}{2t}$ is the stationary phase point for the phase function $\theta=xz + tz^2$. However, if one wants to know the behavior of $u(x,t)$ for moderate values $t$ for which \eqref{eq12} is no longer a good approximation, one must resort to the numerical evaluation of \eqref{eq11}. The problem of
determining $u(x,t)$ for moderate values of $t$ is in two parts. The f\/irst part is the ``forward'' problem and involves the evaluation of the Fourier transform $f\to \hf$. The second part is the ``inverse'' problem and involves the evaluation of \eqref{eq11}. The second part is considerably harder than the f\/irst part because of the high oscillations coming from the multiplier $e^{i (xz +tz^2 )}$. For non-linear integrable systems such as the NLS, KdV, sine-Gordon, \dots, equations, the solution $u(x,t)$ of the equation is expressed in terms of the solution of an associated Riemann--Hilbert problem {RHP} (see, for example, P.~Deift and X.~Zhou~\cite{DZ2003}, and the references therein. S.~Manakov \cite{Manakov} was the f\/irst to observe that the solution of integrable systems (in particular, KdV) could be expressed in terms of a RHP). The associated RHP depends on a product of functions $r(z) e^{\theta_{x,t}(z)}$ where $r(z)$ (the ``ref\/lection'' coef\/f\/icient) encodes the initial data $u_0(x)=u(x,t=0)$ for the equation, and the multiplier $e^{i \theta_{x,t}(z)}$ depends on the equation itself. For example, for NLS $\theta_{x,t}(z) = \theta(z)= xz+tz^2$ as in the case of the free Schr\"{o}dinger equation, and for the KdV equation
\begin{gather*}
u_t + 6 uu_x + u_{xxx} =0 , \qquad -\infty < x < \infty,
\end{gather*}
$\theta_{x,t}(z)= xz + tz^3$, \looseness=1 which is the same multiplier as for the linear KdV equation $u_t + u_{xxx}=0$. We may think of the mapping $r(z) e^{i\theta_{x,t}(z)} \to u(x,t)$ as a generalized Fourier integral transformation, but as opposed to \eqref{eq11}, the transformation is highly nonlinear. As $t\to\infty$, it is possible to use the Deift--Zhou steepest-descent method for RHP's in order obtain an expression for the solution of the equation at hand that is as precise as~\eqref{eq12} in the linear case. For moderate values of~$t$, however, one must again resort to numerical methods. This is a very challenging problem in numerical analysis. Again the problem of the solution of the Cauchy problem is in two parts. The f\/irst part, the forward problem, involves evaluating the ref\/lection map\footnote{In general, one must also append $L^2$ eigenvalues $\{\la_j\}$ and norming constants $(c_j)$ and consider the extended map $u_0\to (r(z), \{\la_j\},\{c_j\} )$.}, $u_0(x) \to r(z)$. This is a spectral/scattering theoretic problem. For example, in the case of KdV, one must solve the spectral/scattering problem for the one-dimensional Schr\"{o}dinger operator $H_0 = -\frac{d^2}{dx^2} + u_0(x)$. The second part, the inverse problem, now involves the numerical solution of a RHP in a situation where the dif\/f\/iculty is again compounded by the high oscillations in the data $r(z) e^{i t \theta_{x,t}(z)}$. The second part is considerably harder than the f\/irst part, and also considerably harder than the evaluation of~\eqref{eq11} in the case of the linear problem. Techniques to solve the forward problem numerically are readily available (see, for example, the recent text~\cite{TObook} and the references therein. In situations in which the forward problem itself contains a small parameter~$\ep$, $\ep \downarrow 0$, there are still many unresolved issues~-- see below). Concerning the second problem, in a very signif\/icant development starting in~2011, S.~Olver~\cite{Olver11} introduced a~methodology to solve RHP's numerically and applied this methodology to compute solutions of the Painlev\'e II equation, which has its own Riemann--Hilbert formulation. This methodology turned out to be suf\/f\/iciently robust to capture the high oscillations arising from factors such as $e^{i\theta_{x,t}(z)}$, and in 2012, B.~Deconinck, S.~Olver and T.~Trogdon~\cite{TOD} used the methodology to solve the Cauchy problem for KdV with remarkable accuracy for remarkably long periods of time, indeed until such time as the steepest-descent method kicks in. In the text~\cite{TObook} of Olver and Trogdon mentioned above, the authors present a framework in which to understand the accuracy of their RHP methodology, and they also discuss applications to other systems.

Rather than a single, or several, open problems, what we have here is an ``\textit{open program}'', viz., to apply the Olver/Deconinck--Olver--Trogdon methodology to solve the many dif\/ferent numerical problems for integrable systems as they arise. One of the most challenging problems concerns situations mentioned above where the forward problem contains a small parameter~$\ep$, $\ep \downarrow 0$ (for KdV, $\ep$~is the small dispersion parameter and for NLS, $\ep= \hbar$, Planck's constant). Such problems have been studied with great interest for more than 30 years, f\/irst by P.~Lax and D.~Levermore \cite{LL1, LL2, LL3}, then by S.~Venakides~\cite{Ven1, Ven2}, then P.~Deift, S.~Venakides and X.~Zhou \cite{VDZ97}, then by P.~Miller and his collaborators \cite{MillerBook, MillerXu}, and many others. Particularly dif\/f\/icult are situations where the forward problem is not self-adjoint, as in the case of focusing NLS. It remains an open problem (see \cite{PercyCM}) to solve the forward problem for focusing NLS with general smooth initial data, and then implement the RHP methodology, to obtain the solution of the Cauchy problem in the small dispersion limit $\ep=\hbar\downarrow 0$. At the technical level, the dif\/f\/iculty in solving the forward problem is that the relevant physical quantities only appear, in the language of Martin Kruskal, ``beyond all orders''. For example, every point $\la$ in the numerical range of the scattering operator $T(\hbar)$ is an eigenvalue of $T(\hbar)$ to all orders, i.e., $\|(T(\hbar) -\la) u\| = \mO (|h|^k )$ for any $k\ge 1$, for some $u=u(k)$, $\|u\|=1$ (see \cite{PercyCM} for more information). The forward problem for focusing NLS should, of course, be seen as one more example of the general open problem of computing the spectrum of non-self-adjoint problems (see, e.g., N.~Trefethen and E.~Embree \cite{Trefethen}, \dots, and particularly E.B.~Davies~\cite{Davies}).

Many problems in modern mathematical physics can be expressed in terms of a Fredholm determinant. For example (see M.~Mehta's classic text \cite{MehtaBook}), $F_s= \det (1-K_s)$ is the pro\-bability that there are no eigenvalues in a~gap of size~$2s/\pi$ in the bulk scaling limit for GUE. Here~$K_s$ has kernel $\frac{\sin s(\la-\mu)}{\pi(\la-\mu)}$, acting on $L^2(-1,1)$. Many authors, starting with F.~Dyson \cite{Dyson62b} in the 1960's, followed by \cite{Dyson76} in the 70's, have contributed to the analysis of~$F_s$ as $s\to\infty$ (see P.~Deift, A.~Its, I.~Krasovsky and X.~Zhou \cite{DIKZ07} for a history of this problem), but in order to evaluate~$F_s$ at f\/inite values of $s$, we must again turn to numerical evaluation. In a~stri\-king development in 2008, F.~Bornemann \cite{Bornemann10} devised a~very general and f\/lexible method to evaluate Fredholm determinants, and he then applied the method to compute a variety of~determinants arising in modern mathematical physics, including~$F_s$. Bornemann's method has emerged as the state of the art for the computation of determinants in RMT and integrable systems. Although Fredholm determinants were introduced more than 100 years, it is surprising that, prior to Bornemann's work, no such general method to evaluate Fredholm determinants numerically and accurately was developed. The operator $K_s$ above on $L^2(-1,1)$ is also of
great interest in many other f\/ields of pure and applied mathematics, in particular in signal processing where one needs to know detailed information about the asymptotic behavior as $s\to\infty$ of the eigenva\-lues~$\{\la_j(s)\}$ and eigenfunctions $\{u_j(s)\}$ of~$K_s$. Quite remarkably $K_s$ commutes with a second-order ordinary dif\/ferential operator, the prolate spheroidal operator, and so the eigenfunctions of~$K_s$ are given by prolate spheroidal functions. The asymptotics for the $\la_j(s)$'s were obtained by Slepian~\cite{Spel} in~1965. The asymptotics of the prolate spheroidal functions are well-known and classical. However, the behavior of the eigenfunctions for f\/inite values of~$s$, a~problem of great practical interest, presents a serious numerical challenge (a~priori, the problem is ill conditioned), which was taken up only very recently by A.~Osipov, V.~Rokhlin and H.~Xiao in their remarkable text~\cite{ORX13}. A~related open problem concerns the Airy operator~$A_s$ with kernel $\frac{A_i(\la) A'_i(\mu)- A'_i(\la) A_i(\mu)}{\la-\mu}$ acting on $L^2(s,\infty)$ (here $A_i(\la)$ is the standard Airy function). Note that $\det (1-A_s)$ is the probability distribution for the largest eigenvalue of a GUE matrix in the edge scaling limit. As in the case of~$K_s$, the asymptotic behavior as $s\to\infty$ of the eigenvalues $\{\tilde{\la}_j(s)\}$ and eigenfunctions $\{\tilde{u}_j (s)\}$ of $A_s$ are of independent interest. For example, it turns out that~$\tilde{\la}_1(s)$ controls the behavior of the solution of the KdV equation in the so-called collisioness shock region. The asymptotics of the $\tilde{\la}_j(s)$'s were derived recently by T.~Bothner~\cite{Bothner16}, conf\/irming an earlier conjecture of C.~Tracy and H.~Widom~\cite{TW}. It is an \textit{open} problem to analyze the behavior of the eigenfunctions $\tilde{u}_j(s)$ as $s\to\infty$, and to compute their behavior for f\/inite~$s$. We note that the $\tilde{u}_j(s)$'s also solve a~second-order ODE (see C.~Tracy and H.~Widom~\cite{TW}).

Finally, in 2005 (see \cite{PercyCM}), the idea was raised of a~``Painlev\'e project'', which would provide a~forum to assemble all current information on the algebraic, analytical and numerical pro\-perties of the Painlev\'e equations. Such a project was indeed launched in the form of a website maintained at NIST, but unfortunately the project has not yet met with much success. Given the importance of the Painlev\'e equations as the core of modern special function theory, such a~project remains a~much needed national resource.
\end{problem}

\begin{problem}[additional problems for integrable systems]\label{prob7}
In 2002, P.~Deift and X.~Zhou \cite{DZ2002} analyzed the behavior of solutions of the Cauchy problem for perturbations of the defocusing NLS equation
\begin{gather}\label{eq13}
\begin{split}
& i u_t + u_{xx}- 2|u|^2 u - \ep |u|^\ell u =0,\\
& u(x,t=0) = f(x) \to 0 \qquad \text{as}\quad |x|\to \infty.
\end{split}
\end{gather}
For $\ell >0$ f\/ixed (for technical reasons they required $\ell > 7/2: \ell >2$ should be enough) they showed, quite surprisingly, that the perturbed equation~\eqref{eq13} remained integrable for $\ep>0$ and suitably small, and they showed that as $t\to \infty$ the solution $u(x,t)=u(x,t; \ep)$ behaved like solutions of NLS (i.e., $\ep=0$). An \textit{open problem} of great and basic interest is to analyze solutions of the Cauchy problem for perturbations
of the focusing NLS equation
\begin{gather}\label{eq14}
\begin{split}
& i\tilde{u}_t+ \tilde{u}_{xx} +2|\tilde{u}|^2 u +\ep |\tilde{u}|^{\ell} \tilde{u} =0,\\
&\tilde{u}(x,t=0) = \tilde{f}(x) \to 0\qquad\text{as}\quad |x|\to \infty.
\end{split}
\end{gather}
Of particular interest is the case where $\tilde{f}(x)$ corresponds to a~2-soliton solution for (unperturbed) focusing NLS. The principal new dif\/f\/iculty in analyzing~\eqref{eq14} over \eqref{eq13}, is that for solutions $u(x,t)$ of \eqref{eq13} we have decay in the sup norm
\begin{gather*}
\sup_x |u(x,t) | \le c/t^{1/2} \to 0 \qquad\text{as}\quad t\to\infty.
\end{gather*}
As $\ell >2$, this makes the perturbation term $\ep|u|^\ell n$ in \eqref{eq13} small compared to $2|u|^2 u$ as $t\to\infty$. Such a decay estimate, however, cannot be true in general for~\eqref{eq14} because of the presence of solitons.

In 1991, S.~Venakides, P.~Deift and R.~Oba \cite{VDO1991} considered the so-called Toda shock problem in which a system of particles $x_k$, $k\ge 0$, interact with exponential forces (cf.\ Problem~\ref{prob5})
\begin{gather}\label{eq15}
\ddx_k = e^{(x_{k-1} - x_k)} - e^{(x_k-x_{k+1})},\qquad k\ge 1
\end{gather}
subject to ``shock'' initial conditions
\begin{gather*}
x_k = k, \qquad \dx_k(0)=0,\quad k\ge 1, \qquad x_0 (t)= 2 at,\quad a>0.
\end{gather*}
The authors showed that for $a> a_{\text{crit}}=1$, in a frame moving with the driving particle $x_0(t)$, the system executes a binary periodic motion as $t\to\infty$. Said dif\/ferently, as $t\to\infty$, the system settles into a 1-gap solution of the doubly inf\/inite periodic Toda lattice. The proof of this result, however, was not fully rigorous. There are two \textit{open} problems here. The \textit{first} is to use Riemann--Hilbert/steepest-descent methods to prove the results in Venakides et al.\ rigorously. The second problem concerns the fact that if we replace the exponential forces~$e^x$ by a general force~$F(x)$,
\begin{gather}\label{eq16}
\ddx_k =F(x_{k-1}- x_k) - F(x_k - x_{k+1}), \qquad k\ge 1
\end{gather}
then numerical simulations show that the solution of the shock problem for this system behaves in the same way as $t\to\infty$ as the Toda system, provided~\eqref{eq16} has a 2-periodic solution
\begin{gather*}
x_k (t+T) = x_k(t), \quad T>0, \qquad x_{k+2}(t) = x_k(t)
\end{gather*}
for $k\in \ZZ$. \textit{Second open} problem: Analyze the general shock problem~\eqref{eq16} for suitably small perturbations~$F(x)$ of~$e^x$.

This paper has been about random matrix theory and integrable systems. Hovering in the background, there is the notion of ``integrable probability theory''. What many people mean by this terminology is that the stochastic problem at hand has special algebraic property which make it possible to solve the system explicitly and ultimately express the key statistics for the problem (e.g., gap probabilities, nearest-neighbor distributions, \dots) in terms of solutions of classically integrable Hamiltonian dynamical systems (e.g., Painlev\'e functions). While this interpretation of the terminology certainly describes accurately the way things work, it ultimately interprets integrable probability theory in terms of a completely dif\/ferent mathematical f\/ield, viz., dynamical systems. One would like, rather, a def\/inition which is intrinsic to probability theory. The salient feature of integrable systems with Hamiltonian $H=H(\xi)$ and $n$ degrees of freedom is that there is a symplectic change of variables $\xi \to (x,y)$ in which the f\/low generated by~$H$ is displayed as $n$ independent and elementary motions
\begin{gather*}
\dx_k=y_k, \qquad \dy_k=0, \qquad 1 \le k\le n,
\end{gather*}
yielding
\begin{gather*}
x_k(t) = x_k(0) + t y_k(0), \qquad y_k(t)=y_k(0), \qquad 1\le k \le n.
\end{gather*}
Now coin f\/lips are certainly elementary probabilistic systems and so a~natural, reasonable and intrinsic probabilistic analog of the dynamical notion of integrability is that there is a change of variables in which key statistics for the stochastic problem at hand become just a product of independent coin f\/lips. And the fact of the matter is that this is precisely the situation for many, if not most, of the known ``integrable probabilistic systems''. For example, for the GUE gap probability $F_s = \det (1-K_s)$ discussed above, the operator $K_s$ is trace-class and satisf\/ies the bounds $0< K_s <1$. Hence, $F_s$ can be expressed as the product of the eigenvalues $\{\la_k(s)\}$ of~$K_s$,
\begin{gather}\label{eq17}
F_s= \prod^\infty_{k=1} (1-\la_k (s)).
\end{gather}
As $0<\la_k(s) <1$, \eqref{eq17} displays the gap probability explicitly as a product of independent f\/lips. Indeed we can imagine that we have a~box and a bag of balls. With probability $\la_1(s)$ the f\/irst ball is placed in the box. With probability $1-\la_1(s)$, it is discarded. With probability $\la_2(s)$, the second ball is placed in the box, and so on.

Similar considerations apply to all determinantal particle systems, with the correlation kernel playing a role similar to the role of a generating function
in integrable Hamiltonian theory. This intrinsic def\/inition of an ``integrable probabilistic system'', is just a post facto observation. I~don't know what
to make of it and whether it will be useful, and I~am just putting it up there for the readers' consideration.
\end{problem}

Finally I would like to raise a problem that has been on my mind for many years. About 120 years ago, in seminal work, A.~Sommerfeld \cite{Sommerfeld} showed how to solve explicitly the dif\/fraction problem for scalar waves at f\/ixed frequency of\/f a semi-inf\/inite slit $\{x>0\}$ in the plane. In the 1930's P.~Morse and P.~Rubenstein \cite{MR} showed how to solve explicitly dif\/fraction from a single f\/inite slit in the plane. However the rigorous solution of the problem of dif\/fraction from two slits remains \textit{open}. This problem is of great mathematical and physical importance, describing as it does Young's famous double slit experiment in optics, and the double slit experiment in quantum mechanics. It is possible that Fokas' methods (see Problem~\ref{prob5}) may be applied to this problem, giving rise to a Riemann--Hilbert solution from which the relevant asymptotics of the solution would easily follow. This is a problem of the
f\/irst order, whose rigorous and explicit solution would generate great interest in the mathematical and physics communities.

We end with an extended paraphrase of \cite{PercyCM}:
\begin{quote}
``There are many other areas, closely related to the problems in the above list, where much progress has been made in recent years, and where much remains to be done. These include: partition functions for random matrix models and their connections to mappings on Riemann surfaces, the representation theory of ``large'' groups such as the inf\/inite symmetric group and the inf\/inite unitary group, Macdonald processes, singular value statistics of matrix products, statistics for normal matrix ensembles, the six vertex model, random matrix models with an external source, non-intersecting random paths and the tacnode phenomenon, SVD statistics, asymptotics for Toeplitz and Hankel matrices, singular values of $n$ products of $m\times m$ random matrices as~$n$, $m\to\infty$, and many others. But as I~said in the beginning, my list is not complete or def\/initive.''
\end{quote}

\subsection*{Acknowledgements}

This work was supported in part by NSF Grant DMS-1300965. The author would
also like to thank Ivan Corwin for his help with Problem~\ref{prob3}, and Jinho Baik and Ivan Corwin for
their unstinting help in preparing this article for publication. The author is also very grateful for the
opportunity to present this list of problems which continue to intrigue and challenge him.

\pdfbookmark[1]{References}{ref}
\LastPageEnding

\end{document}